\documentclass{sig-alternate}

\usepackage{booktabs} % For formal tables

\usepackage{color}
\usepackage{subcaption} 
\usepackage{amsmath}
\usepackage{listings} % code listing

\usepackage{tabularx}
\usepackage{booktabs}

\usepackage{adjustbox}
\usepackage{multirow}

\usepackage{xcolor,colortbl}
\usepackage[inline,shortlabels]{enumitem}

\setlist[itemize]{noitemsep, topsep=0pt}
\setlist[enumerate]{topsep=0pt}

\newcommand{\system}{{\sc Prism}}
\newcommand{\mweaver}{{\sc MWeaver}}
\newcommand{\filter}{{\sc Filter}}

\newcommand{\mondialData}{{\sc Mondial}}

\renewcommand{\theenumii}{\theenumi.\arabic{enumii}.}

\usepackage[linesnumbered,ruled,vlined]{algorithm2e}

\usepackage{hyperref}

\usepackage{pgfplots}
\usetikzlibrary{patterns}
\pgfplotsset{compat=newest}
\usetikzlibrary{decorations.pathmorphing}

\usepackage{listings}

\newcommand{\naive}{na\"{i}ve}

\newcommand{\minisection}[1]{\vspace{0.10cm} \noindent {\bf #1} ---}

\newcommand{\weakdefinition}[1]{{\textbf{\textit{#1}}}. }

\usepackage{tikz}

\newcommand{\eg}{{e.g.}}

\newcommand{\ie}{{i.e.}}

\newcommand{\etal}{{et al.}}
\def\myQuad{\hskip\fontdimen6\font}

\begin{document}
% \copyrightyear{2019} 
% \acmYear{2019} 
% \setcopyright{cidr-2019}
% \acmConference[CIDR'19]{8th Biennial Conference on Innovative Data Systems Research}{Jan 13-16, 2019}{Asilomar, California, USA}
% \acmBooktitle{CIDR'19: 8th Biennial Conference on Innovative Data Systems Research}{Asilomar, California, USA}
% \acmPrice{15.00}
% \acmDOI{10.1145/3209900.3209902}
% \acmISBN{978-1-4503-5827-9/18/06}

\setcopyright{acmcopyright}
\conferenceinfo{CIDR '19}{Janurary 13--16, 2019, Asilomar, CA, USA}

\title{Demonstration of a Multiresolution Schema Mapping System}

\author{Zhongjun Jin\myQuad Christopher Baik\myQuad  Michael Cafarella\myQuad H. V. Jagadish \myQuad Yuze Lou \\
       \affaddr{University of Michigan, Ann Arbor}\\
 \email{\{markjin,cjbaik,michjc,jag,lyzlyz\}@umich.edu}
}
% \affiliation{%
%   \institution{University of Michigan, Ann Arbor}
% %   \streetaddress{P.O. Box 1212}
% %   \city{Dublin}
% %   \state{Ohio}
% %   \postcode{43017-6221}
% }
% \email{{markjin,cjbaik,michjc,jag,lyzlyz}@umich.edu}

\maketitle

\begin{abstract}
Enterprise databases usually contain large and complex schemas. Authoring complete schema mapping queries in this case requires deep knowledge about the source and target schemas and is thereby very challenging to programmers.
Sample-driven schema mapping allows the user to describe the schema mapping using data records. However, real data records are still harder to specify than other useful insights about the desired schema mapping the user might have. In this project, we develop a schema mapping system, \system, that enables {\em multiresolution schema mapping}. The end user is not limited to providing high-resolution constraints like exact data records but may also provide constraints of various resolutions, like incomplete data records, value ranges, and data types. This new interaction paradigm gives the user more flexibility in describing the desired schema mapping. This demonstration showcases how to use \system\ for schema mapping in a real database.
\end{abstract}
% \keywords{Schema mapping, exploratory search, non-expert users, program synthesis, SQL query}

% \setlength{\textfloatsep}{0pt}
\section{Introduction}\label{sec:intro}
Schema mapping is the problem of discovering queries that convert data from source databases with different schemas to a {\em target schema}, \ie, the schema expected by the end user. In real-world complex databases, composing schema mapping queries is challenging because it requires a deep understanding of the source database schema and the target schema. Previous works \cite{qian2012sample,shen2014discovering,Wang2017,mottin2014exemplar,bonifati2017interactive,kalashnikov2018fastqre} have adopted a {\em sample-driven} approach to simplify this process for end users: the user can demonstrate the desired schema mapping process by providing a few data records in the target schema without being familiar with the source database schema. However, we argue that the sample-driven schema mapping approach has two practical challenges:

\begin{enumerate}
\itemsep0em 
\item {\em High-resolution issue}. The user is required to provide {\em high-resolution constraints}, \ie, complete data records with exact values in the target schema. Providing exact values can be challenging for a user unfamiliar with the database content. For example, the user might know the area of Take Tahoe roughly but cannot provide an exact value.
\item {\em Low-expressivity issue}. The user may have insights on the target schema other than data examples, like column names, data types. Existing methods lack mechanisms for capturing these kinds of knowledge.
\end{enumerate}

Take \mondialData---a relational geography data set---as an example. Suppose the goal is to list all lakes, their area and the states they belong to from the \mondialData\ database as in Table~\ref{tab:desiredSchema}. The desired SQL query to obtain such a table is ``\texttt{SELECT geo\_lake.Province, Lake.Name, Lake.Area  
FROM Lake, geo\_lake 
WHERE Lake.Name = geo\_lake.Lake}''. 

A sample-based schema mapping system, such as \mweaver~\cite{qian2012sample}, takes complete target schema data samples from the user and synthesizes schema mapping queries in the form of Project-Join (PJ) SQL queries. A person without the precise geographical knowledge might not be able to use the system because it is hard to specify complete data records in the desired schema, especially the area ({\em High-resolution issue}). However, this does not necessarily mean that the person has no insight to help discover the desired schema mapping query. For example, the person may know that ``Lake Tahoe'' is close to California and Nevada, so one of them must be part of the example. Also, even if the exact lake area of Lake Tahoe is beyond the user's knowledge, she may know that these values must be at least numeric and positive. Although such marginal knowledge should still be useful in narrowing down the search space of possible queries, existing systems cannot use them ({\em Low-expressivity issue}).

\begin{table}[t]
\centering
\small
\begin{tabular}{|l|l|l|}
\hline
 \textbf{State} & \textbf{Lake Name} & \textbf{Area} ($\textit{km}^{2}$) \\ \hline
 California & Lake Tahoe & 497 \\ \hline
 Oregon & Crater Lake& 53.2 \\ \hline
Florida & Fort Peck Lake & 981 \\ \hline
\multicolumn{3}{c}{\dots} \\ 
\end{tabular}
\caption{Desired target schema}
\label{tab:desiredSchema}
\vspace{-.5cm}
\end{table}

% \begin{figure}[t]
% \centering
% \includegraphics[width=.5\linewidth]{image/code.png}
% \caption{Desired schema mapping query}
% \label{fig:fullsql}
% \vspace{-.6cm}
% \end{figure}

\minisection{Our Approach}
To address the above limitations of sample-driven schema mapping, we developed \system\footnote{Available at \url{https://github.com/umich-dbgroup/prism}}, a {\em multiresolution schema mapping} system, that can discover schema mapping queries employing user insights provided at various resolutions.

Multiresolution schema mapping is a schema mapping process with a novel interaction model which increases the scope of descriptions the end user can provide. The model is empowered by a  schema mapping description language enriched to support constraints of various resolutions:
\begin{enumerate*}[1)]
    \item high resolution: complete sample constraints with precise data values,
    \item medium resolution: incomplete sample constraints with approximate data values (a set of possible data values, value ranges),
    \item low resolution: column-level descriptions like data type, value range or even user-defined functions.
\end{enumerate*}

Once the user provides {\em multiresolution constraints} in the proposed language, we synthesize the desired schema mapping query matching these constraints.
A major technical challenge is to ensure that the program search process is efficient enough to be interactive. The search space of all schema mappings is inherently massive; it is exponential in the complexity of the desired schema mapping and the source database schema. Moreover, the number of satisfying solutions can be relatively large because the types of constraints we support are more relaxed than those ingested by the sample-driven approach. As a result, performing a fast search for a complete solution set in our case is difficult. Another challenge is that, for many non-expert users, displaying SQL queries as the output result of a schema mapping system may be difficult to understand. We propose an interactive approach using visualizations to make the synthesized queries more explainable.

In Section~\ref{sec:system}, we present the design of \system. We demonstrate how to use \system\ in Section~\ref{sec:demo}.
\section{System Overview}\label{sec:system}

\begin{figure}[t]
\centering
\begin{equation*}
\begin{split}
\text{Value Constraint }c_{k} :=&\ p_v\ |\ p_v\ logicalop\ p_v\ |\ \epsilon\\
\text{Metadata Constraint }c_{m} :=&\ p_m\ |\ p_m\ logicalop\ p_m\ |\ \epsilon\\
logicalop :=&\   \land\ |\ \lor \\
\text{Value Predicate } p_v :=&\ binop\ const\\
\text{Metadata Predicate } p_m :=&\ type\ binop\ const\\
\text{Metadata Type } type :=&\ \text{DataType}\ |\ \text{ColumnName}\ | \\
&\ \text{MaxValue}\ |\ \text{MinValue}\\
binop :=&\   > | \geq | <  | \leq | = | \neq \\
\end{split}
\end{equation*}
\caption{Multiresolution schema mapping language}
\label{fig:lang}
    \vspace{-.5cm}
\end{figure}

\subsection{Multiresolution Schema Mapping}
\minisection{User Input} 
As enterprise databases today are usually large and complex, users might not have deep understanding about the source database schema or precise knowledge about the database content. In this case, it is difficult for a user to author a schema mapping query or provide even a few data examples in the target schema. However, the user might still have some insights that are useful in narrowing down the space of possible desired schema mapping queries. 

To allow users to comfortably express their insights, extending our previous work~\cite{jin2018beaver}, we propose a \textbf{Multiresolution Schema Mapping Language} (Figure~\ref{fig:lang}), composed of two classes of constraints the users can specify: at the row level, {\em target schema result constraints} (``result constraints'' for short) and, at the column level, {\em target schema metadata constraints} (``metadata constraints'' for short). We also allow the user to add logical operators ``AND'' and ``OR'' between constraint values.

A {\em result constraint} is a set of {\em sample constraints}, which are composed of a sequence of value constraints.
\begin{enumerate}
\itemsep0em 
    \item \weakdefinition{Value Constraints} A value constraint requires a tuple in the target schema to contain a given keyword. Unlike the value constraints handled by traditional schema mapping systems, the user may also specify a disjunction of possible values or a value range. 
    \item \weakdefinition{Sample Constraints} Multiple value constraints listed in the same row together form a sample constraint. A schema mapping query satisfies a sample constraint if the result set of the query contains this sample. Such constraints are also handled by other sample-driven schema mapping systems~\cite{qian2012sample,shen2014discovering}. 
\end{enumerate}

A {\em metadata constraint} represents factual knowledge about individual columns in the source database. Currently, the kinds of metadata we support in \system\ are data type (including \textit{decimal}, \textit{int}, \textit{text}, \textit{date}, \textit{time}), maximum text length, and value range. In the future, we plan to support more metadata constraints, and even user-defined functions. Metadata constraints are allowed to be ``ambiguous'' too: the user could specify a conjunction or disjunction of multiple metadata constraints for one column.

\begin{figure}
    \centering
    \includegraphics[width=\linewidth]{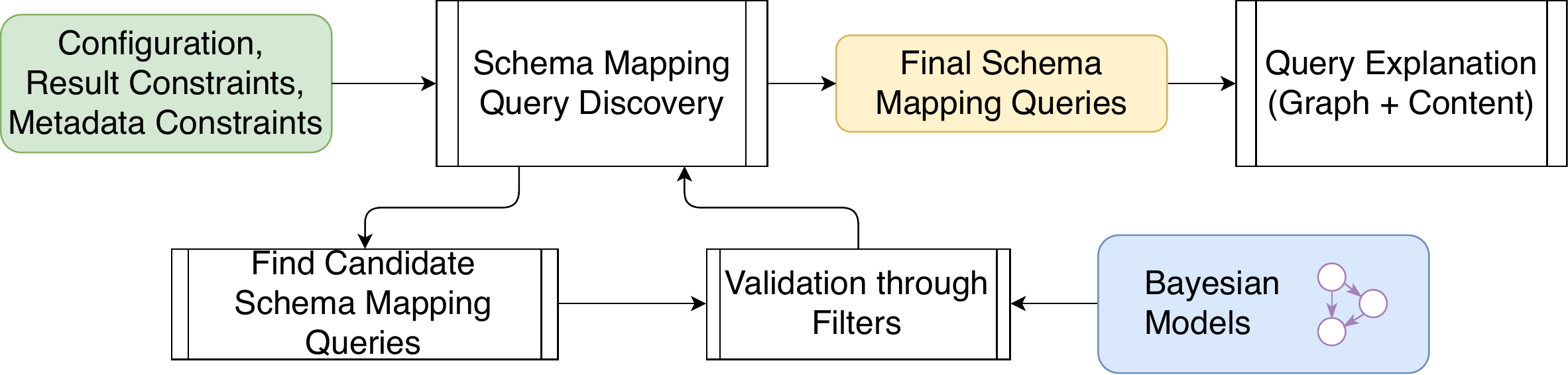}
    \caption{\system\ architecture}
    \label{fig:architecture}
        \vspace{-.3cm}
\end{figure}

\begin{figure}[!t]
\centering
    \centering
      \includegraphics[width=\linewidth]{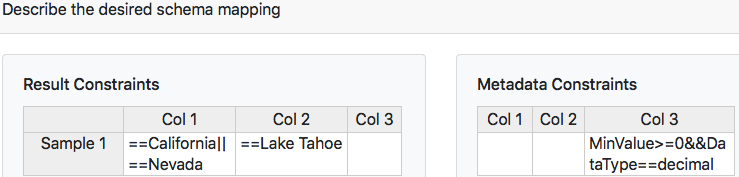}
    \caption{Specify constraints for the desired schema mapping (Description Sec.)}\label{fig:describe}
    \vspace{-.5cm}
    \end{figure}

\minisection{Problem Definition} Given a set of multiresolution schema mapping constraints  $\mathcal{Q}$ (or ``multiresolution constraints'' for short) in the proposed language and a database $\mathcal{D}$, the problem is to synthesize a schema mapping query, $\mathcal{M}$, so that $\mathcal{M}$ and the query result $\mathcal{M}(\mathcal{D})$  satisfy all constraints in $\mathcal{Q}$.

\minisection{System Output}
To focus on the problem without loss of generality, we restrict the space of synthesized schema mapping queries to support Project-Join (PJ) queries. 

\begin{figure*}[!t]
\centering
\begin{subfigure}{.24\textwidth}
\centering
\includegraphics[width=.8\linewidth]{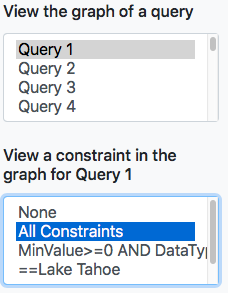}
\caption{Choose to view a synthesized SQL query and its graph}
\label{fig:choose}
\end{subfigure}~
\begin{subfigure}{.22\linewidth}
\centering
\includegraphics[width=.95\linewidth]{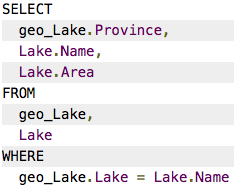}
\caption{Schema mapping query content}
\label{fig:fullsql}
\end{subfigure}~
\begin{subfigure}{.50\linewidth}
\centering
\includegraphics[width=.9\linewidth]{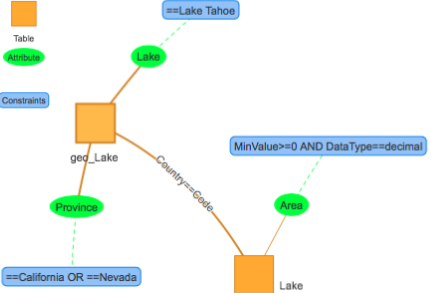}
\caption{The chosen SQL query graph and all constraints}
\label{fig:graph}
\end{subfigure}

\caption{Result Section: show the returned set of schema mapping queries and their graphs}\label{fig:result}
\vspace{-0.5cm}
\end{figure*}

\subsection{System Architecture}
Figure~\ref{fig:architecture} shows \system's architecture and user interaction workflow.
\system\ provides users with a web-based graphical interface with three major sections: \textbf{Configuration}, \textbf{Description} and \textbf{Result}. 

Initially, in the \textbf{Configuration} section, the user sets up the system for the schema mapping task. Current configurations include the source database, number of columns in the target schema, number of sample constraints, and whether metadata constraints are specified.

Next, the user specifies a set of multiresolution constraints, including result constraints and metadata constraints, to describe the desired schema mapping. The \textbf{Description} section (Figure~\ref{fig:describe}) has two regions to take in these constraints.

Once our system obtains the multiresolution constraints, it executes the algorithm introduced in Section~\ref{subsec:algo} to initiate a search for the desired schema mapping query.
In our system, we set a 60-second time limit for each round of query discovery. If \system\ successfully finds a set of schema mapping queries, they are displayed in the \textbf{Result} section (Figure~\ref{fig:result}). If \system\ encounters a timeout, it reports a failure. A synthesized schema mapping query and its results are guaranteed to match the constraints the user initially provided. 

If multiple satisfying schema mapping queries are discovered, the user needs to understand each query and pick the one that is desired. To help the user comprehend each query, in the Result section, \system\ creates a visualization to explain any discovered schema mapping query the user selects (as Figure~\ref{fig:graph}). We discuss this in more detail in Section~\ref{subsec:algo}.

\subsection{Our Approach}\label{subsec:algo}
\minisection{Query Discovery} 
Like \cite{qian2012sample,shen2014discovering}, we split our schema mapping query discovery into of two steps:
\begin{enumerate*}[(\#1)]
    \item discovering candidate complete schema mapping queries, and
    \item validating candidate schema mapping queries.
\end{enumerate*}

The first step, discovering the candidate complete schema mapping queries, is relatively straightforward. First, we identify {\em related columns}---columns in the database potentially used in the schema mapping---so that the search scope for potential schema mapping queries is limited within this small set of columns and tables, and hence, the search space can be significantly reduced. In our setting, finding related columns is essentially finding columns in the database matching at least a value constraint or metadata constraint. Validating sample constraints requires executing expensive join queries on the database and is done in Step 2.

The way we validate a value constraint on a column is same as that in \cite{qian2012sample,shen2014discovering}: leveraging the inverted index provided in most DBMS systems. 
To check a metadata constraint, we use metadata information, \eg, min/max values, collected during preprocessing. With related columns found, we exhaustively search through the source database schema graph and find all possible join paths, each connecting a set of related columns that altogether can be mapped to all columns in the target schema. Every join path along with the set of related columns it connects becomes a {\em candidate} schema mapping query (in form of a PJ query). Note that these candidate schema mapping queries are not final; we have never executed these queries and checked if their query results match the sample constraints.

In Step 2, a \naive\ solution to validating all candidate queries is to execute them one by one on the source database and check their query results, which can be very expensive. In our project, we divide such an expensive verification task into a set of cheap validations of {\em filters}, \ie\ sub(join)trees along with projected attributes (shorter PJ queries), inspired by \cite{shen2014discovering}. 
If a filter fails, its parent filters and entire candidate schema mapping query, from which the filter is derived, automatically fail, and thereby pruned. This gives us an opportunity to replace expensive validations of complex schema mapping queries with cheaper validations of filters.

Although validating filters instead of schema mapping queries saves time, it is still a relatively expensive process. A new important issue becomes the filter validation scheduling: in what order the filters are validated so that the most number of filters are pruned, as well as overall filter validation time is minimized.

A filter scheduling algorithm should naturally consider two important aspects of a filter: pruning power and cost. Estimating the cost of a filter is essentially estimating the cost of executing a SQL query on a database, which is known to be very challenging because the actual cost can be affected by many database tuning parameters, and is beyond the scope of our project. We focus on improving the estimation of pruning power of the filter. 

The pruning power of a filter depends on two things: filter dependency and filter failure (success) probability. While dependency relationships among a set of filters is fixed and can be easily captured, estimating the failure probability of a filter is a bit more tricky. Instead of setting the failure probability proportional to the join path length of a filter~\cite{shen2014discovering}, we take a machine learning approach: we estimate the filter probability using Bayesian models trained a priori for the source database. A Bayesian model is able to give an estimated probability of a certain record matching the sample constraint exists. With this probability and the relation size information, we can obtain a rough estimation of the failure probability good enough to boost our filter scheduling. While learning a Bayesian model in a single relation is no different from learning a model for a data set, learning a model capturing the correlations among multiple relations is more difficult. This problem is solved by using the join indicator introduced by Getoor \etal\ in \cite{getoor2001selectivity}. Details about this idea will be discussed in our future paper.

In the end, we return all final schema mapping queries and let the user choose the desired one.

% \begin{figure}[h]
% \vspace{-.3cm}
% \centering
% \begin{lstlisting}[
%           language=SQL,
%           showspaces=false,
%           basicstyle=\ttfamily,
%           numbers=left,
%           numberstyle=\tiny,
%           commentstyle=\color{gray}
%         ]
% SELECT geo_lake.Province, Lake.Name  
% FROM Lake, geo_lake 
% WHERE Lake.Name = geo_lake.Lake AND
% MATCH (Lake.Name) AGAINST ('"Lake Tahoe"' 
%     IN BOOLEAN MODE) AND
% MATCH (geo_lake.Province) AGAINST ('"California"
%     "Nevada"' IN BOOLEAN MODE) LIMIT 1;
% \end{lstlisting}
% \caption{The actual SQL query for verifying a filter}
% \label{fig:filter}
% \vspace{-.4cm}
% \end{figure}

\minisection{Query Explanation}
In \system, we go beyond simply showing the actual generated SQL queries; we explain the discovered schema mapping queries using visualizations.

Whenever the user points to a schema mapping SQL query (top of Figure~\ref{fig:choose}), we draw a corresponding query graph representation for this query (Figure~\ref{fig:graph}). Orange squares represent relations, green ellipses are the attributes to project, and edges represent join conditions. To help the user understand why a given query matches all the constraints she provides, the user could pick one or more constraints (bottom of Figure~\ref{fig:choose}), and \system\ draws these constraints (as blue boxes) in the previous graph to show the locations in the database where these constraints are satisfied.

\subsection{System Evaluation}
We compared \system\ with \filter\ on a set of synthesized test cases created from a public relational database \mondialData~\cite{may-MONDIAL-report-99}. 
In summary, we observed that the overall execution time of user constraints did not grow significantly as user constraints became loose (containing constraints with disjunctions, value ranges, etc.). Meanwhile, the number of satisfying schema mapping queries discovered did not increase much (unless when there were too many missing values). All these evidences suggest that, \system\ not only requires less user knowledge, it does not increase the user interaction effort in schema mapping. Also, our approach significantly reduced the gap of the required number of filter validations between \filter\ and the optimum (up to $\sim 70\%$; on average $\sim 30\%$), which shows our Bayesian-model-based approach can effectively improve the filter scheduling. This section will be discussed in more details in our future paper.

\section{Demonstration}\label{sec:demo}
Our demonstration aims to show the conference attendees that our proposed system, \system, is able to help a \naive\ user synthesize schema mapping queries using multiresolution constraints.

We use the \mondialData\ data set mentioned in Section~\ref{sec:intro} and two other data sets, IMDB and NBA as the source databases the user can interact with. The user can choose from a set of suggested target schemas or come up with her own. Then, the user is free to provide any multiresolution constraint she can come up with to constrain the desired schema mapping process. In the end, \system\ will show all satisfying schema mapping queries discovered and present visualizations to explain the query the user selects, as discussed in Section~\ref{subsec:algo}.

To best illustrate how to use our tool, we will use the motivating example from Section 1, where a user would do the following\footnote{\url{https://markjin1990.github.io/assets/video/prism.mp4}}:
\begin{enumerate}
\itemsep0em 
    \item In the \textbf{Configuration} section, 1) choose ``\textbf{Mondial}'' as the source database from the supported databases, 2) set the number of columns in the target schema as ``3'', 3) set the number of sample constraints as ``1'', 4) confirm to specify metadata constraints.
    
    \item Specify the multiresolution constraints to describe the desired schema mapping in the \textbf{Description} section.
    
    \begin{enumerate}
    \itemsep0em 
    \item Type ``California || Nevada'' in the first cell in the Sample Constraint field.
    \item Type ``Lake Tahoe'' in the second cell in the Result Constraints field. 
    \item Type ``DataType==`decimal''' AND MinValue>=`0''' in the third cell in the Metadata Constraints field.
    \end{enumerate}
    
    \item Hit the ``Start Searching!'' button.
    \item In the \textbf{Result} section, \system\ shows a list of satisfying schema mapping queries. View the queries and pick the one that is correct.
    \begin{enumerate}
    \itemsep0em 
    \item Select the first synthesized query in the top field in Figure~\ref{fig:choose}. The SQL query is shown as Figure~\ref{fig:fullsql}.
    \item The system draws a graph. 
    \item Interaction: choose the constraints in the bottom field in Figure~\ref{fig:choose} to show in the visualization (Figure~\ref{fig:graph} shows a SQL graph with all constraints user provided). This helps the user understand why the selected query matches the constraints she provided.
    \item If the selected query is not desired, repeat the above process for the second query.
    \end{enumerate}
\end{enumerate}
\section{Conclusion}
Our demonstration shows that the proposed multiresolution schema mapping system, \system, makes schema mapping in a large complex database easy for non-expert users. To build a schema mapping SQL query generating the target schema in a large and complex database, the user may specify constraints of various resolutions, such as disjunctions of values, value ranges, and even column-level metadata constraints which presumably require less domain expertise than high-resolution constraints, \ie, exact data samples. We will continue to develop this system to achieve our vision for a schema mapping system for non-expert users.

\section{ACKNOWLEDGMENTS}
This project is supported in part by by NSF
grants IIS-1250880, IIS-1054913, NSF IGERT grant 0903629, a Sloan Research Fellowship, a CSE Dept. Fellowship, and a University of Michigan MIDAS grant.

\bibliographystyle{plain}
\bibliography{paper}

\end{document}